\documentclass[aps,prd,amsmath
,eqsecnum            
,preprintnumbers
,nofootinbib         
,twocolumn         
]{revtex4}
\usepackage[dvips]{graphicx}
\usepackage{amsmath}
\usepackage{amsfonts}
\usepackage{amssymb}
\usepackage{longtable}   

\allowdisplaybreaks[4]     

\newcommand{\dr}[2]{\frac {{\rm d} {#1}} {{\rm d} {#2}}}

\newcommand\beq{\begin{equation}}
\newcommand\eeq{\end{equation}}
\newcommand\beqa{\begin{eqnarray}}
\newcommand\eeqa{\end{eqnarray}}

\begin{document}

\title{Redshift and redshift-drift in $\Lambda = 0$ quasi-spherical Szekeres cosmological models and the effect of averaging}

\author{Priti Mishra}
\affiliation{Department of Astronomy and Astrophysics, \\
Tata Institute of Fundamental Research, \\
Homi Bhabha Road, Colaba, Mumbai -  400005, Maharashtra, India}
\email{priti@tifr.res.in}

\author{Marie-No\"elle C\'el\'erier}
\affiliation{Laboratoire Univers et Th\'eories (LUTH), \\
Observatoire de Paris, CNRS, Universit\'e Paris-Diderot \\
5 place Jules Janssen, 92190 Meudon, France}
\email{marie-noelle.celerier@obspm.fr}

\date { }

\begin{abstract}
\noindent

Since the advent of the accelerated expanding homogeneous universe model, some other explanations for the 
supernova Ia dimming have been explored, among which there are inhomogeneous models constructed with exact 
$\Lambda = 0$ solutions of Einstein's equations. They have been used either as one patch or to build 
Swiss-cheese models. The most studied ones have been the Lema\^itre-Tolman-Bondi (LTB) models. However,
these models being spatially spherical, they are not well designed to reproduce 
 the large scale structures which exhibit  clusters, filaments and non spherical voids. This is the reason why
Szekeres models, which are devoid of any symmetry, have recently come into play. In this paper, we give the 
equations and an algorithm to compute the redshift-drift for the most general quasi-spherical Szekeres 
(QSS) models with no dark energy. We apply it to a QSS model recently proposed by Bolejko and Sussman 
(BSQSS model) who averaged their model to reproduce the density distribution of 
  the Alexander and collaborators' LTB model which is able to fit a large set of cosmological data without
dark energy. They concluded that their model represents a significant improvement over the
observed cosmic structure description by spherical LTB models. We show here that this QSS model is ruled out by a 
negative cosmological redshift, i.e. a blueshift, which is not observed in the Universe. We also compute
a positive redshift and the redshift-drift for the Alexander et al.'s model and compare this redshift-drift to that
of the $\Lambda$CDM model. We conclude that the process of averaging an unphysical QSS model can lead to
obtain a physical model able to reproduce our observed local Universe with no dark energy need and that
the redshift-drift can discriminate between this model and the $\Lambda$CDM model. For completeness, we also
compute the blueshift-drift of the BSQSS model.
 
\end{abstract}

\maketitle


\section{Introduction} \label{sec1}

In 1998  the SN Ia observations revealed that their observed luminosity was lower than what was expected in the cold 
dark matter (CDM) model \cite{riess98,permutter99}. In other words, the SN Ia were found to be at a distance farther than that predicted by the CDM model. Also, the deceleration parameter was found to be negative in the CDM model. A negative deceleration parameter implies that the Universe expansion rate is accelerating.
This can be explained in FLRW models only if a fluid with negative pressure is assumed to fill the Universe. Such an exotic 
fluid is named dark energy. Since this discovery, there have been many dark energy models proposed in the literature, but none
of them satisfactorily addresses the question of its origin and nature. 

However, there have been attempts to explain these observations without assuming any dark energy component. The main attempts 
can be broadly divided into two categories: inhomogeneous models and modified gravity. As the names suggest, the first category
abandons the space homogeneity assumption and the second category works with modified Einstein's equations (see, e.g., Refs.~\cite{fuzfa06,MS12,moffat06}, and 
Ref.~\cite{TC12} for a review). In this article, we limit ourselves to the study of inhomogeneous models.

The two inhomogeneous solutions of Einstein's equations which have been most frequently used in the literature can be divided
into two classes: Lema\^itre-Tolman-Bondi \cite{GL33,RCT34,HB47} (LTB) models and Szekeres \cite{PS75} models. The LTB metric
is a spatially spherical dust solution of the Einstein equations while the Szekeres metric is a dust solution of these equations
with no symmetry, i.e., no Killing vector \cite{WBB76}. One can find in the literature many LTB models and a few Szekeres models
which claim to explain the cosmological observations without assuming dark energy (see, e.g., Ref. \cite{KB11} for a review and 
also Ref. \cite{AN11} for a study of a particular Szekeres model not included in this previous review). Since these solutions are 
considering only dust as a gravitational source, 
they are valid only in the Universe region where the radiation effect is negligible, i.e., between the last scattering surface 
and our current location. We will use them to study our local Universe where 
dark energy is supposed to have the strongest effect.

In this paper, we are interested in the study of the Szekeres model proposed by Bolejko and Sussman \cite{BS11}, which, once 
spatially averaged reproduces qualitatively the density profile of the Alexander and collaborators' LTB model \cite{ABNV09}. 
This LTB model is a very good fit to the SN Ia data and is also consistent with the WMAP 3-year data and local measurements of 
the Hubble parameter.

However, for models reproducing cosmological data measured on our past light cone, the discrimination between inhomogeneous models 
and $\Lambda$CDM models is impossible. The problem is completely degenerate. This is the reason why several tests using effects
outside the light cone have been proposed, one of these being the source redshift-drift while the observer's proper time is
elapsing \cite{AS62,GMV62}.

In a previous paper \cite{MCS2012}, we have calculated the redshift-drift for the axially symmetric Szekeres model of Ref. \cite{BC10}
and compared it to the redshift-drift in some LTB models found in the literature and to that of the $\Lambda$CDM model. We found that the
redshift-drift is indeed able to distinguish between these different models.

Here, our first purpose was to compute the redshift-drift for the most general Szekeres model of Bolejko and Sussman which displays 
no symmetry and for the LTB Alexander et al.'s model to see whether upon averaging the redshift-drift changes significantly 
and then to compare these redshift-drifts to that in the $\Lambda$CDM model.

We have thus calculated the equations and written a code able to compute, among other features, the redshift and the redshift-drift
of the most general quasi-spherical Szekeres (QSS) model. We have applied this code to the Bolejko and Sussman quasi-spherical Szekeres
(BSQSS) model. We have also computed these quantities for the Alexander et al.'s model with the same recipe used in our previous 
\cite{MCS2012} paper. 
However, we found that the BSQSS model exhibits a negative cosmological redshift, i.e., a blueshift, which is not observed in the
Universe. This must be considered as enough to rule out the model, however, for completeness, we have computed the blueshift-drift
for this model.

The structure of the present paper is as follows. In Sec.~\ref{sec2}, we present the Szekeres models and the particular QSS subclass
used here. In section \ref{sec3} we display the differential equations for the redshift and the redshift-drift in the most general 
QSS models and an algorithm to numerically integrate them. In Sec.~\ref{sec4} we compute the redshift and the redshift-drift in the
model proposed by Bolejko and Sussman \cite{BS11}. In Sec.~\ref{sec5}, we display our results for the redshift and 
redshift-drift computation in the LTB model studied by Alexander et al. \cite{ABNV09}. In Sec.~\ref{sec6}, we present our conclusions.

\section{Szekeres models} \label{sec2}

The Szekeres metric \cite{PS75} is the most general dust solution of Einstein's equations. By the most general solution we mean that this solution has 
no symmetry i.e., it has no Killing vector. In comoving and synchronous coordinates the Szekeres metric is written as

\begin{equation}
{\rm d} s^2 =  c^2 {\rm d} t^2 - {\rm e}^{2 \alpha} {\rm d} r^2 - {\rm e}^{2 \beta} ({\rm d}x^2 + {\rm d}y^2),
\label{metsz}
\end{equation}
where $\alpha \equiv \alpha(t,r,x,y)$ and $\beta \equiv \beta(t,r,x,y) $ are two functions which will be determined by the field equations.

Szekeres solutions are divided into two categories depending upon the value of $\beta'$ where the prime denotes 
derivative with respect to $r$. The class II family, where $\beta' = 0$, is a simultaneous generalization
of the Friedmann and Kantowski-Sachs models \cite{PK06}. Its spherically symmetric limit is the Datt-Ruban
solution \cite{VR68,VR69}. 

The class I family where $\beta'$ is non-zero contains the LTB solution as a spherically symmetric limit.

 Therefore, we
choose this class of solutions to study Szekeres models.  After a change of parameters more convenient for our purpose \cite{H96} and after 
solving Einstein's equations, the class I Szekeres metric can be written as 

\begin{equation}
{\rm d} s^2 =  c^2 {\rm d} t^2 - \frac{(\Phi' - \Phi {  E}'/ {  E})^2}
{\epsilon - k} {\rm d} r^2 - \frac{\Phi^2}{E^2} ({\rm d} x^2 + {\rm d} y^2),
\label{metsz2}
\end{equation}
where $\epsilon = 0, \pm 1$, $\Phi$ is a function of $t$ and $r$, $k$ is a function of $r$, and
\begin{equation}
{  E} = \frac{S}{2} \left[ \left( \frac{x-P}{S} \right)^2
+ \left( \frac{y-Q}{S} \right)^2 + \epsilon \right],
\label{Edef}
\end{equation}
with $S(r)$, $P(r)$, $Q(r)$, functions of $r$.

From (\ref{metsz2}) it can be seen that  all the three Friedmann limits (hyperbolic, flat and spherical) can be achieved only in the case 
where $\epsilon = +1$. This is induced by the requirement of a Lorentzian signature for
the metric. Since we are interested in studying such an inhomogeneous model which becomes homogeneous
at large scales, i.e., before the last-scattering, we consider only the $\epsilon = +1$ case. It is called 
the quasi-spherical Szekeres (QSS) solution. This QSS solution can be imagined as a LTB model generalization in which
the constant mass spheres are non-concentric.

In  the $\epsilon = +1$ case, the Szekeres metric takes the following form:
\begin{equation}
{\rm d} s^2 =  c^2 {\rm d} t^2 - \frac{(\Phi' - \Phi {  E}'/ {  E})^2}
{1 - k} {\rm d} r^2 - \frac{\Phi^2}{E^2}({\rm d} x^2 + {\rm d} y^2),
\label{ds2}
\end{equation}
where
\begin{equation}
{  E} = \frac{S}{2} \left[ \left( \frac{x-P}{S} \right)^2+ \left( \frac{y-Q}{S} \right)^2 +1 \right].
\end{equation}

For the QSS metric (\ref{ds2}) the Einstein equations reduce to the following two:
\begin{equation}
\frac{1}{c^2}\dot{\Phi}^2 = \frac{2M}{\Phi} - k + \frac{1}{3} \Lambda
\Phi^2, \label{vel}
\end{equation}
where the dot denotes derivation with respect to $t$, $\Lambda$ is the cosmological constant and $M(r)$ is an 
arbitrary function of $r$ related to the density $\rho$ via
\begin{equation}
\kappa \rho c^2= 
 \frac{2M' - 6 M {  E}'/{  E}}{\Phi^2 ( \Phi' - \Phi {  E}'/{  E})}, \label{rho}
\end{equation}
where $\kappa=8\pi G/c^4$.

The integration of (\ref{vel}) yields
\begin{equation}
\pm \int\limits_0^{\Phi}\frac{{\rm d} \widetilde{\Phi}}{\sqrt{- k + 2M /
\widetilde{\Phi} + \frac{1}{3} \Lambda \widetilde{\Phi}^2}} = c [ t - t_B(r)].
\label{tbf}
\end{equation}
where $t_B(r)$ is an arbitrary function which is called bang time function and it defines the initial moment of evolution.
When $t_B' \neq 0$, i.e., in general,
this singularity instant is position-dependent, as in the LTB model. The plus sign applies for expanding 
regions. The minus sign applies for collapsing regions. Here, we study the QSS model with $\Lambda = 0$. 

All the equations written so far are covariant under coordinate transformations of the form $\tilde{r} = g(r)$. 
It means that one of the six functions $k(r)$, $S(r)$, $P(r)$, $Q(r)$, $M(r)$ or $t_B(r)$ can be fixed at our 
convenience by the choice of $g$. Hence, each Szekeres solution is fully determined by only five functions of 
$r$ and a coordinate choice. In the BSQSS model, these functions are $S$, $P$, $Q$, $M$ and $t_B$, and the 
coordinate choice is $\Phi(t_{ls}, r) = r$, where the radial coordinate is the areal radius at the last scattering instant, $t_{ls}$.

\section{Redshift and redshift-drift in QSS models} \label{sec3}

As stressed in the introduction, our first aim was to compute the redshift-drift for the BSQSS model and to compare it to
that in the Alexander et al.'s model. However, while calculating this drift, we first computed the redshift and found that it
was negative, i.e., a blueshift. Since a cosmological blueshift is not observed in the Universe, this is enough to rule out 
the BSQSS model as a physical cosmological model. Hence, the redshift-drift, which, in this model, 
would be a blueshift-drift, should also be unphysical. However, we do not claim that such a blueshift is a QSS model general 
feature. Hence, since, to our knowledge, the equations and method to calculate the redshift-drift for the most general QSS models
have never been displayed in the literature, we present them in this section, so that they might be used in future works to 
discriminate between physical QSS models and other models.

\subsection{Definition of the redshift-drift}

\begin{figure}[!htb]
\begin{center} 
 \includegraphics[width=9cm]{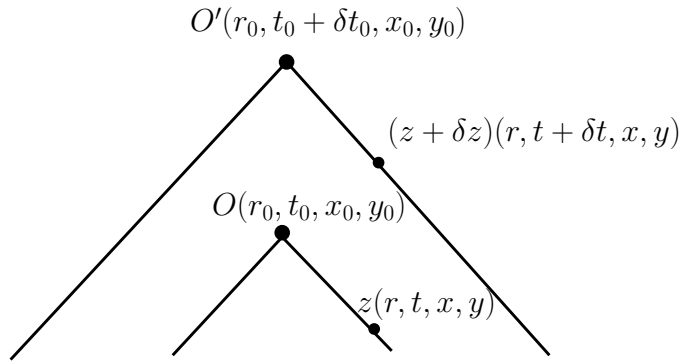}
\caption{The redshift-drift $\delta z$ of a source, initially at a redshift $z$ on the past
light cone of an observer at $O$, as measured by the same observer at $O'$ after an elapsed 
time $\delta t_0$ of the observer's proper time.}
\label{lightcone}
\end{center}
\end{figure}

The redshift-drift is the temporal change in the redshift measured by an observer looking at the same comoving source on her past
light cone at different proper time. It's mathematical definition is $\delta z/\delta t_0$ which is explained schematically in Fig.~\ref{lightcone}.

The redshift-drift has been first calculated by A. Sandage \cite{AS62} and G. McVittie \cite{GMV62} in 1962 for FLRW models. 
Its expression in these models is given by
\beq 
\frac{\delta z}{1+z}=H_0 \delta t_0\left(1-\frac{H(z)}{(1+z)H_0}\right).
\eeq 
Since it is a quantity evolving off the observer's past light cone, it can be used to suppress the degeneracy between models 
reproducing the same cosmological data on this light cone. It has been calculated for some LTB models \cite{QA10,Yoo11}, an 
axially symmetric QSS model \cite{MCS2012}, Stephani models \cite{dabrowski13a}, and in varying speed of light (VSL) theory 
\cite{dabrowski13b} with no dark energy. Since axially symmetric QSS models \cite{BC10} are not realistic universe models, 
we give  two new recipes for calculating the redshift-drift for QSS models which do not exhibit any symmetry.

\subsection{The equations for the redshift and the redshift-drift}

The geodesic equations for the QSS model in $(t,r,x,y)$ coordinates are \cite{BKHC09}

\beqa
&& c^2\frac {{\rm d}^2 t} {{\rm d} s^2} + \frac {\Phi,_{tr} - {\Phi,_t} {
E},_r/{E}} {1 - k} (\Phi,_r - \Phi {E},_r/{E})
\left(\dr r s\right)^2 \nonumber \\
&& + \frac {\Phi \Phi,_t} {{E}^2} \left[\left(\dr x s\right)^2  + \left(\dr
y s\right)^2\right] = 0, \label{tge} \\
&& \frac {{\rm d}^2 r} {{\rm d} s^2} + 2 c \frac {{\Phi,_{tr}} - {\Phi,_t}{E},_r / { E}} {\Phi,_r - \Phi { E},_r/{ E}} \dr t s \dr r s \nonumber \\
&+& \left(\frac {\Phi,_{rr} - \Phi { E},_{rr}/{ E}} {\Phi,_r - \Phi
{ E},_r/{ E}} - \frac {{ E},_r} { E} + \frac 1 2 \frac {k,_r}
{1- k}\right) \left(\dr r s\right)^2 \nonumber \\
&+& 2 \frac {\Phi} {{ E}^2} \frac {{ E},_r  { E},_x - { E} {
E},_{xr}} {\Phi,_r - \Phi { E},_r/{ E}} \dr r s \dr x s \nonumber \\
&+& 2 \frac {\Phi} {{ E}^2} \frac {({ E},_r { E},_y - { E} {E},_{yr})} {\Phi,_r - \Phi {E},_r/{ E}} \dr r s \dr y s \nonumber \\
&-& \frac {\Phi} {{ E}^2} \frac {1 - k} {\Phi,_r - \Phi {E},_r/{ E}} \left[\left(\dr x s\right)^2 + \left(\dr y s\right)^2\right] =
0, \nonumber \\ \label{rge} \\
&& \frac {{\rm d}^2 x} {{\rm d} s^2} + 2 c \frac {\Phi,_t} {\Phi} \dr t s \dr x s
\nonumber \\
&-& \frac 1 {\Phi} \frac {{\Phi},_r - {\Phi} { E},_r/{ E}} {1- k} ({ E},_r { E},_x - { E}  { E},_{xr})
\left(\dr r s\right)^2 \nonumber \\
&+& \frac 2 {\Phi} \left(\Phi,_r - \Phi \frac {{ E},_r} { E}\right) \dr
r s \dr x s - \frac {{ E},_x} { E} \left(\dr x s\right)^2 \nonumber \\
&-& 2 \frac {{ E},_y} { E} \dr x s \dr y s + \frac {{ E},_x} {E} \left(\dr y s\right)^2 = 0, \label{xge} \\
&& \frac {{\rm d}^2 y} {{\rm d} s^2} + 2 c \frac {\Phi,_t} {\Phi} \dr t s \dr y
s \nonumber \\
&-& \frac 1 {\Phi} \frac {\Phi,_r - \Phi {E},_r/{ E}} {1 - k}
({ E},_r { E},_y - { E} { E},_{yr}) \left(\dr r s\right)^2
\nonumber \\
&+& \frac 2 {\Phi} \left(\Phi,_r - \Phi \frac {{ E},_r} { E}\right) \dr
r s \dr y s + \frac {{ E},_y} { E} \left(\dr x s\right)^2 \nonumber \\
&-& 2 \frac {{ E},_x} { E} \dr x s \dr y s - \frac {{ E},_y} {E} \left(\dr y s\right)^2 = 0. \label{yge}
\eeqa 
And the null condition is
\beqa 
&&c^2\left(\frac{\rm{d}t}{\rm{d}s}\right)^2-\frac{\left(\Phi_{,r}-\Phi E_{,r}/E\right)^2}{1-k}\left(\frac{\rm{d}r}{\rm{d}s}\right)^2
-\frac{\Phi^2}{E^2}\left(\left(\frac{\rm{d}x}{\rm{d}s}\right)^2\right. \nonumber \\
&&\left.+\left(\frac{\rm{d}y}{\rm{d}s}\right)^2\right)=0.
\label{ncond}
\eeqa

Now, we choose the $r$ coordinate as the affine parameter, using the following transformation relation:
\beq 
\dr {^2 x^\mu} {s^2} = \left(\dr r s\right)^2 \dr {^2 x^\mu} {r^2} + \dr{^2 r} {s^2} \dr {x^\mu} r.\label{transfeqn}
\eeq
Then, from (\ref{rge}) we have
\beqa
&& \frac {{\rm d}^2 r} {{\rm d} s^2} = \left(\dr r s\right)^2 \left\{- 2c\frac
{\Phi_{01}} {\Phi_1} \dr t r - \left(\frac{\Phi_{11}} {\Phi_1} - \frac {
E_{,_r}} { E} + \frac 1 2 \frac {k,_r} {1 - k}\right)\right.
\nonumber \\
&& - \left.2 \frac{\Phi}{{ E}^2} \frac{E_{12}} {\Phi_1} \dr x r - 2 \frac
{\Phi} {{ E}^2} \frac {E_{13}} {\Phi_1 } \dr y r + \frac {\Phi} {{ E}^2}
\frac {1 - k} {\Phi_1} \Sigma\right\} \nonumber \\
&& = U(t, r, x, y) \left(\dr r s\right)^2.
\eeqa

where \beqa
\Phi_{1}&=&\Phi_{,r}-\Phi E_{,r}/E,\\
\Phi_{01}&=&\Phi_{,tr}-\Phi_{,t} E_{,r}/E,\\
\Phi_{11}&=&\Phi_{,rr}-\Phi E_{,rr}/E,\\
E_{12}&=& E_{,r} E_{,x}- EE_{,xr},\\
E_{13}&=&E_{,r} E_{,y}- EE_{,yr},\\
\Sigma&=&\left(\frac{\rm{d}x}{\rm{d}r}\right)^2+\left(\frac{\rm{d}y}{\rm{d}r}\right)^2
\eeqa
and 
\beqa 
&& U=-2c\frac{\Phi_{01}}{\Phi_{1}}\frac{\rm{d}t}{\rm{d}r}-\frac{\Phi_{11}}{\Phi_{1}}+\frac{E_{,r}}{E}-\frac{1}{2}\frac{k_{,r}}{1-k}
\nonumber \\
&& -2\frac{\Phi}{E^2}\frac{E_{12}}{\Phi_{1}}\frac{\rm{d}x}{\rm{d}r}-2\frac{\Phi}{E^2}\frac{E_{13}}{\Phi_{1}}\frac{\rm{d}y}{\rm{d}r}
+\frac{\Phi}{E^2}\frac{1-k}{\Phi_{1}}\Sigma.
\eeqa

The above transformation will bring the geodesic equations and the null condition equation in the following form:
\beqa 
&& c^2\frac {{\rm d}^2 t} {{\rm d} r^2}+\frac{\Phi_1\Phi_{01}}{1-k}+\frac{\Phi\Phi_{,t}}{E^2}\Sigma+cU\frac{\rm{d}t}{\rm{d}r} =0\label{teqn}\\
&&\frac {{\rm d}^2 x} {{\rm d} r^2}+2c\frac{\Phi_{,t}}{\Phi}\frac{\rm{d}t}{\rm{d}r}\frac{\rm{d}x}{\rm{d}r}-
\frac{1}{\Phi}\frac{\Phi_1}{1-k}E_{12}+\frac{2\Phi_1}{\Phi}\frac{\rm{d}x}{\rm{d}r}
-\nonumber \\
&&\frac{E_{,x}}{E}\left(\frac{\rm{d}x}{\rm{d}r}\right)^2-2\frac{E_{,y}}{E}\frac{\rm{d}x}{\rm{d}r}\frac{\rm{d}y}{\rm{d}r}
+\frac{E_{,x}}{E}\left(\frac{\rm{d}y}{\rm{d}r}\right)^2+\nonumber \\
&&U\frac{\rm{d}x}{\rm{d}r}=0\label{xeqn}\\
&&\frac {{\rm d}^2 y} {{\rm d} r^2}+2c\frac{\Phi_{,t}}{\Phi}\frac{\rm{d}t}{\rm{d}r}\frac{\rm{d}y}{\rm{d}r}-
\frac{1}{\Phi}\frac{\Phi_1}{1-k}E_{13}+\frac{2\Phi_1}{\Phi}\frac{\rm{d}y}{\rm{d}r}
+\nonumber \\
&&\frac{E_{,y}}{E}\left(\frac{\rm{d}x}{\rm{d}r}\right)^2-2\frac{E_{,x}}{E}\frac{\rm{d}x}{\rm{d}r}\frac{\rm{d}y}{\rm{d}r}
-\frac{E_{,y}}{E}\left(\frac{\rm{d}y}{\rm{d}r}\right)^2+\nonumber \\
&&U\frac{\rm{d}y}{\rm{d}r}=0\label{yeqn}
\eeqa
 
\beq 
c^2\left(\frac{\rm{d}t}{\rm{d}r}\right)^2-\frac{\Phi_1^2}{1-k}-\frac{\Phi^2}{E^2}\Sigma=0.\label{nulleqn}
\eeq

The equation for the redshift in QSS models is \cite{PK06,BKHC09}
\beq
\frac{{\rm d}z}{{\rm d}r} = \frac{1 + z}{c} \frac{\dot{\Phi}' - \dot{\Phi}E'/E}{\sqrt{1 - k}}.
\label{dzdreqn}
\eeq

Initially the observer's coordinates are $(t(s_o),r(s_o),x(s_o),y(s_o))$, which we write $(t_0,r_0,x_0,y_0)$, and 
the source coordinates are $(t(s_e),r(s_e),x(s_e),y(s_e))$, which we write $(t_e,r_e,x_e,y_e)$.  

Substituting $t=t+\delta t$ in (\ref{nulleqn}), and keeping terms only up to first order in $\delta t$, we get  
\beq 
c^2\left(\frac{\rm{d}(t+\delta t)}{\rm{d}r}\right)^2-\frac{(\Phi_1+\dot{\Phi}_1\delta t)^2}{1-k}
-\frac{(\Phi + \dot{ \Phi} \delta t)^2}{E^2}\Sigma=0.
\label{tplusdeltat}
\eeq 
Now subtracting (\ref{nulleqn}) from (\ref{tplusdeltat}), and still keeping terms only up to first order in $\delta t$, we get
\beq 
c^2\frac{\rm{d} t}{\rm{d}r}\frac{\rm{d}\delta t}{\rm{d}r}-\frac{\Phi_1\dot{\Phi}_1}{1-k}\delta t
-\frac{\Phi \dot{ \Phi} }{E^2}\delta t\Sigma=0.\label{deltateqn}
\eeq 
Substituting $z=z+\delta z$ and $t=t+\delta t$ in (\ref{dzdreqn}) we obtain

\beq
\frac{{\rm d}(z+\delta z)}{{\rm d}r} = \frac{1 + z+\delta z}{c} \frac{\dot{\Phi}'(t+\delta t) 
- \dot{\Phi}(t+\delta t)E'/E}{\sqrt{1 - k}}.
\label{zplusdeltaz}
\eeq
 
Subtracting (\ref{dzdreqn}) from (\ref{zplusdeltaz}), and keeping terms only up to first order  in $\delta t$ and $\delta z$, we get
\beq
\frac{{\rm d}(\delta z)}{{\rm d}r} = \frac{1 +z}{c\sqrt{1 - k}} \dot{\Phi}_{01}\delta t+\frac{\delta z}{c\sqrt{1 - k}}\Phi_{01}.
\label{deltazeqnqss}
\eeq 
We will solve (\ref{deltazeqnqss}) together with (\ref{teqn})-(\ref{yeqn}) and (\ref{deltateqn}) to get the redshift-drift in
QSS models.

The redshift-drift can also be calculated by the following method.

Initially the observer's coordinates are $(t(s_o),r(s_o),x(s_o),y(s_o))$, which we write $(t_0,r_0,x_0,y_0)$, and 
the source coordinates are $(t(s_e),r(s_e),x(s_e),y(s_e))$, which we write $(t_e,r_e,x_e,y_e)$. 
The redshift of this source is $z$ given by
\beqa 
1+z&=&\frac{k_t^{e}}{k_t^{o}},\\
&=&\frac{{\rm d}t/{\rm d}s|_{s=s_e}}{{\rm d}t/{\rm d}s|_{s=s_o}}.\label{zeqn}
\eeqa 
After some proper time elapse $\delta t_0$ at the observer's location, the observer's coordinates become
$(t_0 + \delta t_0,r_0,x_0,y_0)$, and the source coordinates become $(t_e+\delta t(s_e),r_e,x_e,y_e)$. Since 
we are working with comoving coordinates, $r$, $x$ and $y$ do not change.

Substituting $t=t+\delta t$ in (\ref{tge})-(\ref{yge}), we get,
\beqa
&&c^2\frac{{\rm d}^2 \left(t+\delta t\right)}{{\rm d} s^2} + 
\left(\frac{{\Phi,_{tr}} - {\Phi,_t} {E},_r/{E}}{1 - k} (\Phi,_r - \Phi {E},_r/{E})\right)  \nonumber \\
&&\left(t+\delta t,r,x,y\right)  \left( \frac{{\rm
d} r}{{\rm d} s} \right)^2  + \left(\frac{\Phi {\Phi,_t}}{{E}^2}\right) \left(t+\delta t,r,x,y\right) \nonumber \\
&& \left[ \left(
\frac{{\rm d} x}{{\rm d} s} \right)^2
 + \left( \frac{{\rm d} y}{{\rm d} s} \right)^2 \right] = 0,\label{tplusdeltateqn}
\eeqa
\beqa
&&\frac {{\rm d}^2 r} {{\rm d} s^2} + 2 c \left(\frac {{\Phi,_{tr}} - {\Phi,_t}{E},_r/{E}}{\Phi,_r - \Phi {E},_r/{E}}\right)
 \left(t+\delta t,r,x,y\right)
\frac{{\rm d} \left(t+\delta t\right)}{{\rm d} s}
\frac{{\rm d} r}{{\rm d} s} 
\nonumber \\
&&+ \left(\left( \frac{\Phi,_{rr} - \Phi,_r {E},_r/{E} - \Phi {
E},_{rr}/{E} + \Phi ({E},_r/{E})^2}{\Phi,_r - \Phi {
E},_r/{E}}\right) \right. \nonumber \\
&& \left. \left(t+\delta t,r,x,y\right) + \frac{1}{2}\frac{k,_r}{1-k} \right) \left( \frac{{\rm d}r}{{\rm d} s} \right)^2 
\nonumber \\
&&+ 2 \left(\frac{\Phi}{{E}^2} \frac{{E},_r  {E},_x - {E}
 {E},_{xr}}{\Phi,_r - \Phi {E},_r/{E}} \right)\left(t+\delta t,r,x,y\right)\frac{{\rm d}
r}{{\rm d} s} \frac{{\rm d} x}{{\rm d} s} 
\nonumber \\
&&+ 2 \left(\frac{\Phi}{{E}^2}
\frac{({E},_r {E},_y - {E}  {
E},_{yr})}{\Phi,_r - \Phi {E},_r/{E} } \right)\left(t+\delta t,r,x,y\right)\frac{{\rm d} r}{{\rm d} s}
\frac{{\rm d} y}{{\rm d} s} 
\nonumber \\
&&- \left(\frac{\Phi}{{E}^2} \frac{1-k}{\Phi,_r - \Phi {E},_r/{E}}\right)\left(t+\delta t,r,x,y\right)
\left[ \left( \frac{{\rm d} x}{{\rm d} s} \right)^2 \right. \nonumber \\
&& \left. + \left( \frac{{\rm d}
y}{{\rm d} s} \right)^2 \right] = 0,
\label{rplusdeltareqn}
\eeqa
\beqa
&&\frac{{\rm d}^2 x}{{\rm d} s^2} + 2 c \frac{{\Phi,_t}}{\Phi} \left(t+\delta t,r\right)\frac{{\rm d}
\left(t+\delta t\right)}{{\rm d} s}  \frac{{\rm d} x}{{\rm d} s} - \left( \frac{1}{\Phi}
\frac{{\Phi},_r - {\Phi} {E},_r/{E}}{1 - k} \right. \nonumber \\
&& \left. ({E},_r  {E},_x
- {E}  {E},_{xr}) \right)\left(t+\delta t,r,x,y\right) \left(\frac{{\rm d} r}{{\rm d} s} \right)^2
\nonumber \\
&&+ 2 \left(  \frac{\Phi,_r}{\Phi}\left(t+\delta t,r\right) - \frac{{E},_r}{{E}} \right)
\frac{{\rm d} r}{{\rm d} s}  \frac{{\rm d} x}{{\rm d} s} -
 \frac{ {E},_x}{{E}} \left( \frac{{\rm d} x}{{\rm d} s}
 \right)^2 \nonumber \\
&& - 2 \frac{{E},_y}{{E}}
\frac{{\rm d} x}{{\rm d} s}  \frac{{\rm d} y}{{\rm d} s} 
+ \frac{ {E},_x}{{E}} \left( \frac{{\rm d} y}{{\rm d} s} \right)^2 =
0, 
\label{xplusdeltaxeqn}
\eeqa
\beqa 
&&\frac{{\rm d}^2 y}{{\rm d} s^2} + 2 c \frac{{\Phi,_t}}{\Phi}\left(t+\delta t,r\right) \frac{{\rm d}
\left(t+\delta t\right)}{{\rm d} s}  \frac{{\rm d} y}{{\rm d} s} - \left(\frac{1}{\Phi} \frac{\Phi,_r
- {\Phi} {E},_r/{E}}{1 - k} \right. \nonumber \\
&& \left. ({E},_r  {E},_y - {E}
{E},_{yr}) \right)\left(t+\delta t,r,x,y\right) \left( \frac{{\rm d} r}{{\rm d} s} \right)^2
\nonumber \\
&&+ 2 \left( \frac{\Phi,_r}{\Phi}\left(t+\delta t,r\right) - \frac{{E},_r}{{E}} \right)
\frac{{\rm d} r}{{\rm d} s}  \frac{{\rm d} y}{{\rm d} s} + \frac{ {
E},_y}{{E}} \left( \frac{{\rm d} x}{{\rm d} s} \right)^2 \nonumber \\
&& - 2 \frac{{
E},_x}{{E}} \frac{{\rm d} x}{{\rm d} s} \frac{{\rm d} y}{{\rm d} s}
- \frac{{E},_y}{{E}} \left( \frac{{\rm d} y}{{\rm d} s} \right)^2 =0. 
\label{yplusdeltayeqn}
\eeqa
Subtracting (\ref{tge}) from (\ref{tplusdeltateqn}), and keeping only the first order terms in $\delta t$, we obtain
\beqa 
&& c^2 \frac{{\rm d}^2 \delta t}{{\rm d} s^2} + 
\frac{\partial}{\partial t}\left(\frac{{\Phi,_{tr}} - {\Phi,_t} {E},_r/{E}}{1 - k} \left(\Phi,_r - \Phi \frac{{E},_r}{{E}}\right)\right)\delta t \left( \frac{{\rm
d} r}{{\rm d} s} \right)^2\nonumber \\
&&  + \frac{\partial}{\partial t}\left(\frac{\Phi {\Phi,_t}}{{E}^2}\right) \delta t \left[ \left(
\frac{{\rm d} x}{{\rm d} s} \right)^2
 + \left( \frac{{\rm d} y}{{\rm d} s} \right)^2 \right] = 0
\label{deltateqn_s}
\eeqa
Now we want to calculate the change in redshift $\delta z$ which would be observed after a proper time elapse $\delta t_0$ at the observer's location. 
We proceed as follows.

The new redshift $(z+\delta z)$ is given by
\beq 
1+z+\delta z=\frac{{\rm d}\left(t+\delta t\right)/{\rm d}s|_{s=s_e}}{{\rm d}\left(t+\delta t\right)/{\rm d}s|_{s=s_o}}
\label{zplusdeltazeqn}
\eeq
Subtracting (\ref{zeqn}) from (\ref{zplusdeltazeqn}), we obtain
\beqa 
\delta z&=&\frac{{\rm d}\left(t+\delta t\right)/{\rm d}s|_{s=s_e}}{{\rm d}\left(t+\delta t\right)/{\rm d}s|_{s=s_o}}
-\frac{{\rm d}t/{\rm d}s|_{s=s_e}}{{\rm d}t/{\rm d}s|_{s=s_o}}\\
\delta z &=&\frac{\left({\rm d}t/{\rm d}s+{\rm d}\delta t/{\rm d}s\right)|_{s=s_e}}{\left({\rm d}t/{\rm d}s+{\rm d}\delta t/{\rm d}s\right)|_{s=s_o}}
-\frac{{\rm d}t/{\rm d}s|_{s=s_e}}{{\rm d}t/{\rm d}s|_{s=s_o}}\\
\delta z &=&\frac{{\rm d}t/{\rm d}s|_{s=s_e}}{{\rm d}t/{\rm d}s|_{s=s_o}}\left(\frac{1+\frac{{\rm d}\delta t/{\rm d}s}{{\rm d}t/{\rm d}s}|_{s=s_e}}
{1+\frac{{\rm d}\delta t/{\rm d}s}{{\rm d}t/{\rm d}s}|_{s=s_o}}\right) \nonumber \\
&-&\frac{{\rm d}t/{\rm d}s|_{s=s_e}}{{\rm d}t/{\rm d}s|_{s=s_o}}\\
\delta z&=&(1+z)\left(\frac{\frac{{\rm d}\delta t/{\rm d}s}{{\rm d}t/{\rm d}s}|_{s=s_e}-\frac{{\rm d}\delta t/{\rm d}s}{{\rm d}t/{\rm d}s}|_{s=s_o}}
{1+\frac{{\rm d}\delta t/{\rm d}s}{{\rm d}t/{\rm d}s}|_{s=s_o}}\right)
\label{deltazeqn}
\eeqa 

We also write equation (\ref{deltazeqn}) for the redshift-drift as
\beq
\delta z=(1+z)\left(\frac{\frac{{\rm d}\delta t/{\rm d}r}{{\rm d}t/{\rm d}r}|_{r=r_e}-\frac{{\rm d}\delta t/{\rm d}r}{{\rm d}t/{\rm d}r}|_{r=r_o}}
{1+\frac{{\rm d}\delta t/{\rm d}r}{{\rm d}t/{\rm d}r}|_{r=r_o}}\right)
\label{deltaz}
\eeq

This is another equation to calculate the redshift-drift in any QSS model. By solving (\ref{deltaz}) one can compute the change in redshift of
a source at $(t_e,r_e,x_e,y_e)$ after a $\delta t_0$ proper time has elapsed at the observer's initial location in space-time, $(t_o,r_o,x_o,y_o)$. 
The quantities ${\rm d}\delta t/{\rm d}r$ and ${\rm d}t/{\rm d}r$ appearing in (\ref{deltaz}) are obtained by solving (\ref{teqn})-(\ref{nulleqn}) simultaneously.

\subsection{Calculation of the function k(r)} \label{kr}

Since we know $M(r)$, the function $k(r)$ is needed to compute $\Phi$ from the parametric solution of (\ref{vel}) 
we can obtain when $\Lambda=0$, once we have determined the sign of $k(r)$. There are two different methods for
calculating $k(r)$ depending on the value of the $t_B(r)$ function.

\subsubsection{$t_B(r) \neq 0$}

In this case, we use the parametric method. However, there are three different parametric solutions depending on the sign of $k(r)$.

\begin{enumerate}

\item $k>0$

\beq
\Phi(t,r) = \frac{M}{k} (1 - \cos\eta),
\label{solkpos1}
\eeq
and
\beq
t-t_B(r)= \frac{M}{k^{3/2}} (\eta-\sin \eta).
\label{solkpos2}
\eeq

\item $k=0$

\beq
\Phi(t,r) = \left[\frac{9}{2} M \left(t-t_B(r)\right)^2 \right]^{1/3}.
\label{solknull}
\eeq

\item $k<0$

\beq 
\Phi(r,t)=\frac{M}{(-k)}(\cosh\eta-1),
\label{solutionforeeq1}
\eeq 
and 
\beq  
t-t_B(r)=\frac{M}{(-k)^{3/2}}(\sinh\eta-\eta),
\label{solutionforeeq2}
\eeq 
where $\eta(t,r)$ is the parameter.

\end{enumerate}

We do not know a priori what is the $k(r)$ sign. Therefore, we have to try the above three solutions at random. 
Our coordinate choice is $\Phi(t_0, r)=r$. This choice helps us to determine the function $k(r)$ as following.

The case $k=0$ is the easiest to deal with. Setting $t=t_0$ in (\ref{solknull}) and replacing $M(r)$ and $t_B(r)$ by
their expressions, and $\Phi(t_0, r)$ by $r$, we see at once whether (\ref{solknull}) is fulfilled for some given $r$
values. There might be indeed cases when $k$ vanishes for some $r$ value(s) and $k$ changes sign (or not) at this (these) value(s). 
In this cases, we have to test $k<0$ and $k>0$ for the different $r$ ranges, between the values where $k$ is null. 
If $k$ vanishes nowhere, we just guess the sign of $k$ for all the $r$ values and proceed as follows.

We give here the reasoning for $k<0$. Since at $t=t_0$, $\Phi(t_0, r)=r$ and $\eta(t_0,r) = \eta_0(r)$, we set $t=t_0$ 
in (\ref{solutionforeeq1}) and (\ref{solutionforeeq2}) and eliminate $k(r)$ between both. We obtain
\beq
-k = \frac{M}{r}(\cosh \eta_0 - 1)
\label{cosheta0}
\eeq
\beq
t_0-t_B(r)=\frac{r^{3/2}}{M^{1/2}}\frac{(\sinh\eta_0 - \eta_0)}{(\cosh \eta_0 -1)^{3/2}}
\label{rooteq}
\eeq
We keep the non-vanishing root of (\ref{rooteq}) for $\eta_0(r)$ and we substitute it in (\ref{solutionforeeq1})
where we have set $t=t_0$ to get $k(r)$.

An analogous method applies for the case $k>0$.

\subsubsection{$t_B(r)=0$}

In this case, we do not need to guess a priori the sign of $k(r)$. It proceeds directly from the calculations. 
However, we must guess the sign in front of the integral in (\ref{tbf}), since we do not know whether the region of 
the model we are considering is expanding or collapsing. Since we are supposed to study a cosmological model, we 
could guess that the plus sign applies, but we will see in the following that the BSQSS
model region of interest is blueshifted and therefore collapsing.

As an example, we describe this method with the plus sign. The method with the minus sign follows easily. We set 
$t=t_0$ and $\Lambda=t_B(r)=0$ in (\ref{tbf}) with the plus sign and obtain
\beq
\int_0^r\frac{{\rm d} \widetilde{\Phi}}{\sqrt{- k(r) + 2M(r) /
\widetilde{\Phi}}} = c t_0.
\label{intt01}
\eeq

To avoid divergences due to the $1/\widetilde{\Phi}$ term, we multiply the integrand by $\sqrt{\widetilde{\Phi}}$. Eq.~(\ref{intt01}) becomes
\beq
\int_0^r\sqrt{\frac{\widetilde{\Phi}}{- k(r)\widetilde{\Phi} + 2M(r)}}{\rm d} \widetilde{\Phi} = c t_0.
\label{intt02}
\eeq

Now, for a given $r$ value,

\begin{itemize}

\item We choose a $k(r)$ value in this function definition interval, i.e., $-\infty < k < 1$. We span 
this interval with $k$ values separated by some given step. Since we cannot span all this interval towards negative
values, we begin with taking as limits $-1<k<1$ (if necessary, we try an extended interval afterwards). We try first
$k=-1$, then $k=1$, since we will use an interpolation method to find $k$.

\item We insert each $k$ value, and that of $M$ for the given $r$ value, in the integral of (\ref{intt02}) which 
we integrate with, e.g., the trapezium method, with a $r/n$ integration step. We write
$\int_0^r\sqrt{\frac{\widetilde{\Phi}}{- k\widetilde{\Phi} + 2M}}{\rm d} \widetilde{\Phi}=\sum\limits_{i=1}^n 
\sqrt{\frac{\widetilde{\Phi}_i}{- k\widetilde{\Phi}_i + 2M}}\delta \widetilde{\Phi} = I(r),$
with $\delta \widetilde{\Phi}=r/n$ and $\widetilde{\Phi}_i=ir/n$.

\item Then, we check whether $I(r)=ct_0$. If this is the case, that means that the $k$ value chosen corresponds 
actually to the given $r$ value. If not, we try another value for $k$ using an interpolation method and so on. 
By this method, we are able to check whether the interval $-1<k<1$ is satisfactory or whether we need to extend it towards more negative values. 

\end{itemize}

We reiterate the above calculation for a number of $r$ values spanning the light cone section of interest.

\subsection{The Algorithm}

In order to calculate the redshift and the redshift-drift, we proceed in the following manner:

\begin{enumerate}

\item Once $k(r)$ is determined by one of the above methods, we use the corresponding parametric solution for 
$\Phi$ to find $\Phi(t,r)$ and its derivatives on the past light cone.

\item We substitute the $\Sigma$ value from (\ref{nulleqn}) into the geodesic equations (\ref{teqn})-(\ref{yeqn})
to transform them into null geodesic equations.

\item Then we split the three second order null geodesic equations thus obtained into six first order ordinary equations. 
 
\item We find $t(r)$, $x(r)$, $y(r)$ and their first order derivatives on the past light cone by numerically solving
these null geodesic equations. The initial conditions at the current observer where $r=r_o$ are chosen as $t=t_0$, $x=x_0$,
$y=y_0$, ${\rm d}x/{\rm d}r={\rm d}x/{\rm d}r|_0$ 
and ${\rm d}y/{\rm d}r={\rm d}y/{\rm d}r|_0$, and the initial condition for ${\rm d}t/{\rm d}r$ is determined from the null condition.

\item Then we find the redshift $z$ by numerically integrating (\ref{zeqn}).

\item After having found $z$, we find $\delta t$ and $\delta z$ by numerically solving (\ref{deltateqn}) and (\ref{deltazeqnqss}) 
together.

\item Then we find the redshift-drift $\delta z/\delta t_0$ with $\delta t_0=10$ yrs.

\end{enumerate}

\section{Computation of the redshift and the redshift-drift in the BSQSS model} \label{sec4}

\subsection{The BSQSS Model}

The BSQSS model is defined at the last scattering surface by specifying five among its six arbitrary functions of 
$r$ and one coordinate choice. The five functions are $t_B(r), M(r), S(r), P(r), Q(r)$.

We have chosen this model because on spatial averaging, it has been shown in Ref.~\cite{BS11} that the averaged model
reproduces qualitatively the MV model of Ref.~\cite{ABNV09} which fits SN Ia and WMAP data and is consistent with the local $H_0$ value. 

In the BSQSS model, the bang time function, $t_B(r)$, is null and the $M(r)$ function is given by
\[ M(r) = 4 \pi \frac{G}{c^2} \int_0^r 
 \rho_b(1+ \delta\bar{\rho})\, \bar r^2\,{\rm d} \bar r,\]
where $\delta \bar{\rho}=-0.005{\rm e}^{-(\ell/100)^2}+ 0.0008 {\rm e}^{-[(\ell-50)/35]^2}
+0.0005 {\rm e}^{-[(\ell-115)/60]^2}	
+ 0.0002 {\rm e}^{-[(\ell-140)/55]^2}$,
and $\ell \equiv r/$ 1 kpc.

The functions $Q, P,$ and $S$ are defined as follows

$S=1 \Rightarrow S' = 0,$

${\cal D} = 1.05 (1+r)^{-0.99} {\rm e}^{-0.004 r},$

$Q' = {\cal D}$, $P'=0$ for $\ell \leqslant 27,$

$Q' = -{\cal D}$, $P'=0$ for $27 < \ell \leqslant 35,$

$Q' = 0$, $P'=- {\cal D}$ for $35 < \ell \leqslant 41,$

$Q' = 0$, $P'= {\cal D}$ for $41 < \ell \leqslant 51.5,$

$Q' = 0.88 {\cal D} $, $P'=- 0.5 {\cal D}$ for $51.5 < \ell \leqslant 61,$

$Q' = 0.71 {\cal D} $, $P'=0.71 {\cal D}$ for $61 < \ell \leqslant 69,$

$Q' = 0 $, $P'=- {\cal D}$ for $69 < \ell \leqslant 77,$

$Q' = - {\cal D} $, $P'= 0$ for $77 < \ell \leqslant 86.5,$

$Q' = 0.74 {\cal D} $, $P'=-0.74 {\cal D}$ for $86.5 < \ell \leqslant 96,$

$Q' ={\cal D} $, $P'= {\cal D}$ for $96 < \ell \leqslant 102,$

$Q' = - {\cal D} $, $P'= 0$ for $102 < \ell \leqslant 115,$

$Q' = {\cal D} $, $P'= 0$ for $115 < \ell \leqslant 129,$

$Q' = 0 $, $P'=- {\cal D}$ for $\ell >129.$

\subsection{Calculation of the function k(r)}

Since $t_B(r)=0$, we could have used the second method described in Sec.~\ref{kr} to compute $k(r)$. However, 
we faced a problem in our numerical 
calculations since an a priori expanding cosmological model was not compatible with the BSQSS model. Of course, 
we could have changed the sign in (\ref{intt01}), but we found that the first method used less CPU time, since 
the same equations give $k(r)$ and $\Phi(t,r)$.

\begin{figure}[!htb]
\begin{center} 
 \includegraphics[width=9cm]{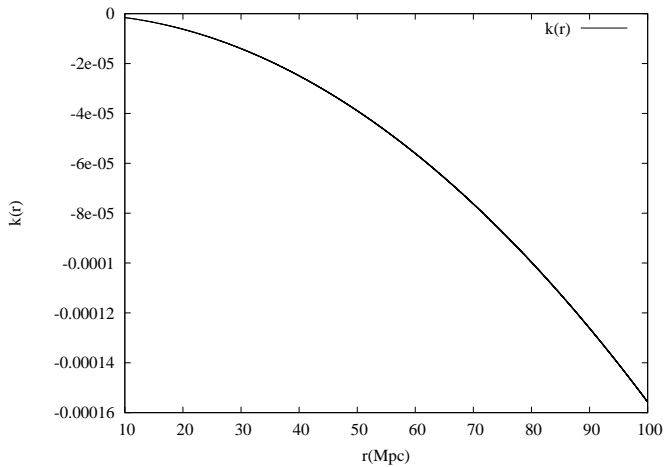}
\caption{The $k(r)$ function in the BSQSS model.}
\label{kofr}
\end{center}
\end{figure}

Therefore, we switched to the first method and found that the equations with $k<0$ for all $r$ gave us a proper 
solution to our problem. The $k(r)$ function is displayed in Fig.~\ref{kofr}.

\subsection{The redshift}

We have runned our code over a huge number of initial conditions, and we have always found the same qualitative results for the redshift, in particular its sign.

\begin{figure}[!htb]
\begin{center}
 \includegraphics[width=9cm]{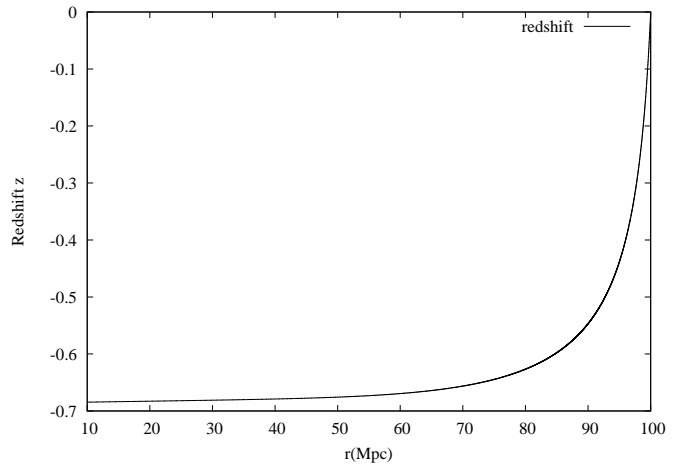}
\caption{The negative redshift (blueshift) as a function of the comoving distance $r$ for the BSQSS model.}
\label{zvr_bsqss}
\end{center}
\end{figure}

For Fig.~\ref{zvr_bsqss} displayed here, the initial conditions at the current observer where $r=r_o=100$ Mpc are
\beqa 
t=t_o=13.7 \, \text{Gyr},\\
x=0.00001,\\
y=0.00001,\\
{\rm d}x/{\rm d}r=0.0001,\\
{\rm d}y/{\rm d}r=0.0001.
\eeqa
and the initial condition for ${\rm d}t/{\rm d}r$ is determined from the null condition.

Fig.~\ref{zvr_bsqss} shows the redshift as a function of the comoving distance $r$ in the BSQSS model. This redshift is 
found to be negative which means that the light rays reaching the observer in a BSQSS universe are blueshifted. We observe
this blueshift because the observer's location is not at this model origin which is at the last scattering surface. 
Since in this model the universe is expanding away from this origin, the sources are coming towards the observer which is at 
$t=t_0$ and $r_0=100$ Mpc. Hence, the light rays are blueshifted. Since such a cosmological blueshift is not observed in the 
Universe, this means that the non averaged BSQSS model is ruled out as a cosmological model.

\subsection{The redshift-drift (blueshift-drift)}

\begin{figure}[!htb]
\begin{center}
 \includegraphics[width=9cm]{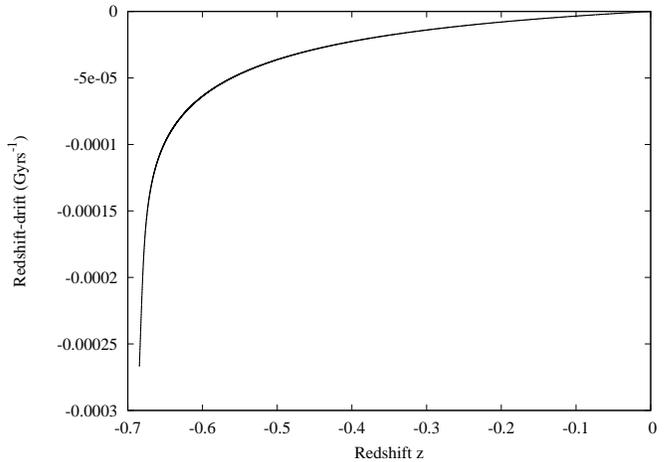}
\caption{The redshift-drift (blueshift-drift) as a function of the negative redshift (blueshift) $z$ for the BSQSS model.}
\label{redshift_drift_BSQSS}
\end{center}
\end{figure} 

However, for completeness, we have used the above described recipe to compute the redshift-drift 
(which is actually a blueshift-drift) in this model. The result is depicted in Fig.~\ref{redshift_drift_BSQSS} 
where we have plotted the blueshift as a negative redshift. For this calculation, we have set the proper 
time elapse $\delta t_0$ to the value of 10 years, i.e., $10^{-8}$ Gyr. This is the reason why the redshift-drift is 
given in units Gyr$^{-1}$. We see from Fig.~\ref{redshift_drift_BSQSS} that the redshift-drift is negative and that 
its effect is very small, of order yrs$^{-13}$ at a blueshift of around 0.7. Of course, since the
model is already ruled out by the blueshift, we do not need to worry about measuring such a small drift, but this 
computation shows that our recipe and our code for calculating the redshift-drift work well and can be used for other 
general Szekeres models with no symmetry.

\section{Averaging effect: the redshift and the redshift-drift in the MV model} \label{sec5}

It has been shown in Ref.~\cite{BS11} that, once spatially averaged, the BSQSS model reproduces qualitatively the density
profile of the LTB MV model of Ref.\cite{ABNV09} with a central observer. We calculate in this section the MV model redshift
to see what becomes of the BSQSS blueshift once the model is averaged. We find this blueshift becomes a cosmological redshift
and then, to discriminate it from the $\Lambda$CDM model, we compute the MV model redshift-drift.

\subsection{LTB models}

LTB models are spatially spherically symmetric solutions of Einstein's equations with dust as a gravitational source. Their
metric in comoving and synchronous time gauge is, with the usual notations \cite{BKHC09}
\beq
{\rm d} s^2 = - c^2 {\rm d} t^2 + \frac{R'^2}{1 + 2E(r)}{\rm d} r^2 + R^2(t,r) ({\rm d} \theta^2 + \sin^2 \theta {\rm d} \phi^2),
\label{ltbmetric}
\eeq
where $E(r)$ is an arbitrary function (corresponding to $-k(r)/2$ in QSS models and to $\bar{M}r^2 k(r)$ in the MV model) and 
$R(t,r)$ obeys the same equation (\ref{vel}) as $\Phi(t,r)$ in QSS models, in which a new arbitrary function of $r$, $M(r)$, 
appears. A third arbitrary function, the $t_B(r)$ bang time, appears as an integration constant of (\ref{vel}) in (\ref{tbf}).
Hence, an LTB solution can be defined by three arbitrary functions of $r$, $E(r)$, $M(r)$ and $t_B(r)$.

The mass density in energy units is
\beq
\kappa \rho = \frac{2M'}{R'R^2},
\label{rholtb} 
\eeq
with $\kappa=8 \pi G/c^4$.

In the MV model, the cosmological constant $\Lambda$ is also set to zero, since the aim is to reproduce the cosmological 
observations without dark energy. Then, the solutions to (\ref{vel}) are the same as (\ref{solkpos1})-(\ref{solutionforeeq2}),
with an inverse sign for $E$ as regards the one for $k$ in the QSS models.

\subsection{The equation for the redshift-drift}

After averaging the BSQSS model, the current observer is located at the center of the occurring LTB model \cite{BS11}. 
Therefore, we give below the redshift-drift equation for a central observer.

\begin{figure}[!htb]
 \begin{center}
 \includegraphics[width=9cm]{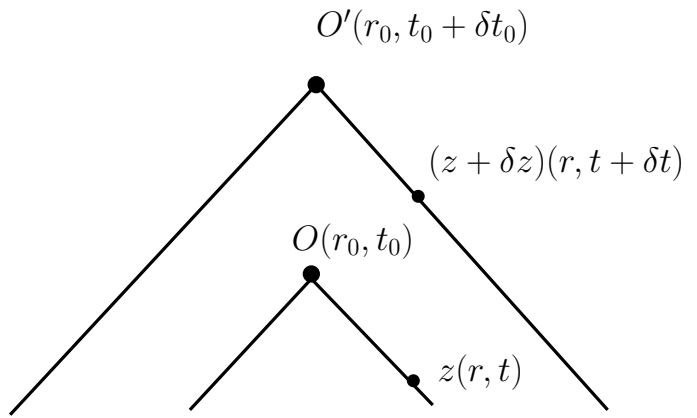}
\caption{The redshift-drift $\delta z/\delta t_0$ of a source, initially at a redshift $z$, measured by the same observer 
at $O$ and $O'$, in a LTB model.}
\end{center}
\end{figure}
We consider a comoving observer $O$ located at the origin, with coordinates $(t_0, r=0)$. The observer receives the light
emitted by a comoving source at $(t,r)$. We denote this source redshift by $z(t,r)$. After a $\delta t_0$ proper 
time elapse, the comoving observer moves to a new location, $O'$ $(t_0 + \delta t_0, r=0)$ and the comoving source moves 
to the new coordinates $(t+\delta t,r)$. Now, this source redshift observed at $O'$ will be 
\begin{equation}
Z(r) = z(r) + \delta z(r),
\label{eq1}
\end{equation}
and its time coordinate
\begin{equation}
T(r) = t(r) + \delta t(r),
\label{eq2}
\end{equation}
with $t(r=0) = t_0$, $z(r=0) = Z(r=0) = 0$, $\delta z(r=0) = 0$ and $\delta t(r=0) = \delta t_0$.

The equation for the redshift is
\begin{equation}
\frac{{\rm d}z}{{\rm d}r} = \frac{1 + z}{c} \frac{\dot{R}' }{\sqrt{1 + 2E}}.
\label{eq3}
\end{equation}

Differentiating (\ref{eq1}) with respect to $r$ and re-arranging the terms, it comes 
\begin{equation}
\frac{{\rm d}\delta z(r)}{{\rm d}r} = \frac{{\rm d}Z(r)}{{\rm d}r} - \frac{{\rm d}z(r)}{{\rm d}r}.
\label{eq5}
\end{equation}

Using (\ref{eq3}) in (\ref{eq5}) and keeping only the first order terms in $\delta z$ and $\delta t$ since they are 
very small compared to $z$ and $t$, we obtain
\begin{eqnarray}
\frac{{\rm d}\delta z}{{\rm d}r} &=& \frac{1 + z}{c} \frac{\ddot{R}'(t,r)}{\sqrt{1 + 2E}} \delta t \\ \nonumber
&+& \frac{\dot{R}'(t,r) }{\sqrt{1 + 2E}} \frac{\delta z}{c}.
\label{eq8}
\end{eqnarray}

Differentiating (\ref{eq2}) with respect to $r$ and re-arranging the terms, it comes 
\begin{equation}
\frac{{\rm d}\delta t(r)}{{\rm d}r} = \frac{{\rm d}T(r)}{{\rm d}r} - \frac{{\rm d}t(r)}{{\rm d}r}.
\label{eq9}
\end{equation}

Using the null condition equation, with the minus sign for incoming light rays, in (\ref{eq9}) and keeping only the first
order term in $\delta t$, we obtain
\begin{equation}
\frac{{\rm d}\delta t(r)}{{\rm d}r} = - \frac{1}{c} \frac{\dot{R}'(t,r) }{\sqrt{1 + 2E}} \, \delta t.
\label{eq11}
\end{equation}

We consider the case where the redshift $z$ is monotonically increasing with $r$. We replace the independent variable $r$
by $z$ by using the following chain rule of differentiation: 
\begin{equation}
\frac{{\rm d}}{{\rm d}r} = \frac{{\rm d}z}{{\rm d}r} \frac{{\rm d}}{{\rm d}z} = 
\frac{1 + z}{c} \frac{\dot{R}'}{\sqrt{1 + 2E}} \frac{{\rm d}}{{\rm d}z}.
\label{eq12}
\end{equation}

Using (\ref{eq12}) in (\ref{eq8}) and re-arranging the terms, we obtain
\begin{equation}
\frac{{\rm d}\delta z}{{\rm d}z} = \frac{\ddot{R}' }{\dot{R}'} \delta t + \frac{\delta z}{1 + z}.
\label{eq14}
\end{equation}

Similarly, using the transformation equation (\ref{eq12}) in (\ref{eq11}) and re-arranging the terms, we obtain 
\begin{equation}
\frac{{\rm d}\delta t}{{\rm d}z} = - \frac{\delta t}{1 + z}.
\label{eq16}
\end{equation}

We integrate (\ref{eq16}) from the observer O at $(t_0, z=0)$ to the source at $(t, z)$ and obtain
\begin{equation}
\delta t = \frac{\delta t_0}{1 + z}.
\label{eq17}
\end{equation}

We insert this expression for $\delta t$ into (\ref{eq14}) and obtain the equation for the redshift-drift:
\begin{equation}
\frac{{\rm d}}{{\rm d}z} \left(\frac{\delta z}{1 + z}\right) = \frac{1}{(1 + z)^2} \frac{\ddot{R}'}{\dot{R}' } \delta t_0.
\label{eq19}
\end{equation}
We numerically integrate (\ref{eq19}) for a fixed $\delta t_0$ value to obtain $\delta z$ and then we calculate the 
redshift-drift from its definition $\dot{z}=\delta z / \delta t_0$.

\subsection{The MV Model}

This MV model is a void, in an Einstein-de Sitter (EdS) background, with minimal under-density contrast around -0.4, 
and minimal radius of order 200 - 250 Mpc/$h$ able to reproduce the SN Ia data with no dark energy and to be consistent
with the 3-yr WMAP data and measurements of the local Hubble parameter $H_0$.

In this LTB void model the mass function, the curvature function and the bang time function are defined as follows, in
units $c=G=1$ and the Planck mass obeying $M_p^2=8 \pi$,
\beq 
M(r)=\frac{1}{6}\bar{M}^2 M_p^2 r^3,
\label{mrmv}
\eeq
\beq
E(r)=(\bar{M}r)^2k_{max}\left[1-\left(\frac{r}{L}\right)^4\right]^2,
\label{ermv}
\eeq
\beq
t_B(r)=0
\label{tbrmv}
\eeq
where $\bar{M}$, $k_{max}$ and $L$ are parameters of the model and $E(r)$ is positive or null. 

The $\bar{M}$ parameter is an arbitrary unphysical mass scale, related to the Hubble parameter via the following relation: 
\beq 
\bar{M} = \sqrt{\frac{3}{8\pi}}\frac{h_{out}}{3000},
\label{mbarmv} 
\eeq
where $h_{out}$ is the Hubble parameter in the EdS region. 
 
One can see from (\ref{ermv}) that the $k_{max}$ parameter corresponds to the amplitude of the density fluctuation inside
the void and $L$ is the void radius beyond which the universe is described by a flat EdS metric.

For this model best fit to the data, the parameter values are $h_{out}=0.452$, $k_{max}=5.302$ and $L$ is 250 Mpc/$h$ where $h=.55$.

\subsection{The algorithm for the MV model}

In our numerical calculations, we use units in which the fundamental constants are set to their usual values. Notice that
the factor 1/3000 in (\ref{mbarmv}) appears for a $1/c$ factor. In order to calculate the redshift and the redshift-drift, 
we proceed as follows.

\begin{enumerate}

\item First, we compute $t(r)$ on the past light cone by numerically solving the following null condition equation for
incoming geodesics in LTB models:
\beq 
 \frac{{\rm d} t}{{\rm d} r} = - \frac{1}{c} \frac{R'}{\sqrt{1 + 2E}}.
\label{nullconditionltb}
\eeq
  
\item Since $E(r)$, corresponding to the quantity we denoted $-k(r)/2$ in QSS models, is nearly everywhere positive, we 
use the parametric solution for QSS $k(r)$ negative. Substituting $t(r)$ in (\ref{solutionforeeq2}) we obtain $\eta(r)$, 
using which in (\ref{solutionforeeq1}) we calculate $R(t(r),r)$ and its derivatives on the past light cone.
\footnote{The parametric equations are the same for QSS and LTB models because (\ref{vel}) with a vanishing $\Lambda$ is the same.}

\item Then we numerically solve the following equation for the redshift $z(t(r),r)$
\beq 
 \frac{{\rm d} z}{{\rm d} r} = \frac{1+z}{c} \frac{\dot{R}'}{\sqrt{1+2E}}.
\eeq

\item After having found $z$, we compute the redshift-drift at this $z$ by numerically solving (\ref{eq19}).

\end{enumerate}

\subsection{The Results}

\begin{figure}[!htb]
\begin{center}
 \includegraphics[width=9cm]{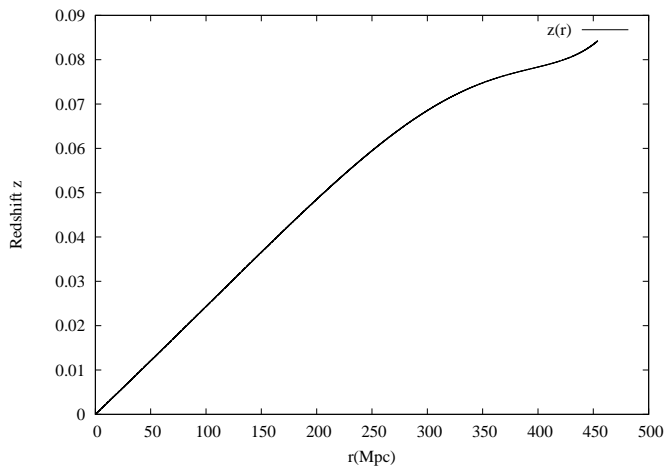}
\caption{The redshift ($z$) as a function of the comoving radial coordinate $r$ for the MV model.}
\label{zmv}
\end{center}
\end{figure}

Fig.~\ref{zmv} shows the redshift in the MV model up to the border of the void where $r=L=450$ Mpc and $z=0.085$. It is quite proportionally 
increasing with $r$ up to around 300 Mpc above which it exhibits a strange feature. This might be due to a non proper matching between the 
void and the background EdS universe.

\begin{figure}[!htb]
\begin{center}
 \includegraphics[width=9cm]{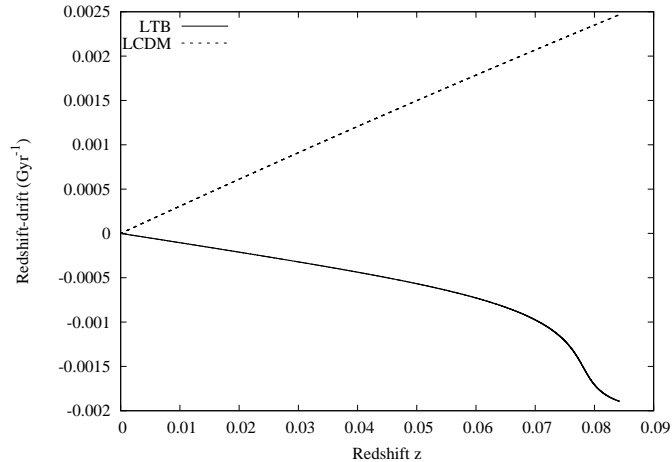}
\caption{The redshift-drift ($\delta z / \delta t_0$) as a function of the redshift $z$ for the MV model and the $\Lambda$CDM model.}
\label{drift_alex_lcdm}
\end{center}
\end{figure}

Fig.~\ref{drift_alex_lcdm} depicts the redshift-drift behavior as a function of the redshift in the MV LTB model and in the $\Lambda$CDM model. 
In the redshift range of interest, the redshift-drift in this LTB model remains negative while in the $\Lambda$CDM model it is positive. In principle, 
this may allow us to discriminate between both models even if they reproduce the same observational data. Also, in both models, the magnitude of the 
redshift-drift increases monotonically with the redshift. However, it is a very small effect, of order  Tyr$^{-1}$ at the void border in the MV model, and therefore, very difficult to observe in future experiments.

\section{Conclusions} \label{sec6}

The type  Ia supernova data, when analyzed in a FLRW framework, seems to be revealing that  our Universe expansion is accelerating from redshifts that 
correspond to non-linear structure formation. In the standard $\Lambda$CDM cosmological model, this is put down to the effect of a dark energy component
which, up to now, is not understood. Among different other explanations, the use of exact inhomogeneous models with no dark energy to reproduce 
the cosmological data has been rather extended in the literature. The first models used have been of the LTB class. These are dust spherically 
symmetric models and have been used either to build one patch models or to construct Swiss-cheese models (see, e.g., Ref. \cite{KB11} for a review).
However, we observe that the structures in the Universe are not spherically symmetric. Therefore, $\Lambda=0$ Szekeres models with no symmetry are
now coming into play (see, e.g., \cite{BC10,KB11,AN11,BS11,MCS2012}), the ones most frequently used being of the quasi-spherical class \cite{BKHC09}.

Now, these Szekeres models are much more complicated to deal with and the first authors who used them as cosmological models added some symmetry,
e.g., axial \cite{BC10,MCS2012}. Then, other studies have been made with Szekeres models with no symmetries \cite{BS11,IPT13}. However, it is very tricky to
reproduce directly cosmological data with such models.

This is the reason why, in Ref.~\cite{BS11}, the authors have considered a very general quasi-spherical Szekeres model, then spatially averaged it and obtained
the LTB MV model density profile of Ref.~\cite{ABNV09}. Since this MV model reproduces the SN Ia data and is consistent with the 3-yr WMAP data and the local 
Hubble parameter measurements, the Szekeres model of Ref.~\cite{BS11} can be considered as a proper inhomogeneous  model which,
once coarse-grained and averaged,
is consistent with these data set. This strengthens the argument proposed in Ref.~\cite{MNC12} that void model spherical symmetry is but a mathematical 
simplification of an energy density smoothed out over angles around us.

Now, models which reproduce the same cosmological data as $\Lambda$CDM ones on the observer past light cone cannot be distinguished from this model. 
The problem is completely degenerate. This is the reason why we have been interested in calculating the redshift-drift of both models with a view to
comparing them, first between them, then to that of the $\Lambda$CDM model. We have therefore, for the first time in the literature to our knowledge,
given two equation sets and an algorithm to compute the redshift-drift in the most general QSS model. Then, we have applied them to the BSQSS model of Ref.~\cite{BS11}. 
One of the steps to obtain the redshift-drift is to calculate the redshift and, in doing this for the BSQSS model, we have found that this redshift was negative,
i.e., a blueshift. We observe this blueshift because the observer's location is not at the origin in this model. Actually, the origin is at the
last scattering surface. Since, in this model, the universe is expanding away from this origin, the sources 
are coming towards the observer which is at $t=t_0$ and $r_0=100$ Mpc. Hence, the light rays are blueshifted. Since such a cosmological blueshift
is not observed in the Universe, this means that the non averaged BSQSS model is ruled out as a cosmological model. However, we cannot claim it 
is a generic feature of all quasi-spherical Szekeres models.

However, for completeness, and to test our recipe and our code, we have calculated the redshift-drift (blueshift-drift) for the BSQSS model. 
We have found that this redshift-drift is negative, that its amplitude is increasing with the blueshift and that it is a very tiny effect. Indeed, for 
a ten year observation, and around a blueshift of $z=-0.7$, the blueshift variation amplitude is $|\delta z| \sim 10^{-12}$. However, since the 
model is already ruled out by its blueshift, the redshift-drift consideration is purely theoretical.

It has been shown in Ref.~\cite{BS11} that, once spatially averaged, the BSQSS model reproduces qualitatively the density profile of the LTB MV 
model of Ref.\cite{ABNV09} with a central observer. We have thus calculated the MV model redshift to see what becomes of the BSQSS blueshift once 
the model is averaged. We have found that this blueshift becomes a cosmological redshift and then, to discriminate it from the $\Lambda$CDM model,
we have computed the MV model redshift-drift. This redshift appeared to be negative, with an amplitude increasing with redshift. On the contrary, 
in the redshift range of interest, the $\Lambda$CDM model redshift-drift is positive which, in principle, would allow one to discriminate between
both models by measuring their drift. However, these redshift-drifts are also very tiny effects, since the void border is only at a small redshift of $z \sim 0.085$.
At this redshift, the redshift variation amplitude of the MV model, for a ten year observation, is merely $|\delta z| \sim 2.10^{-11}$.
This will not be measurable by the future experiments dedicated to the redshift-drift measurement in the Universe like CODEX/EXPRESSO \cite{codex07,JL08,QA10} 
and the gravitational waves observations DECIGO/BBO \cite{YNY12}.

However, the model proposed in Ref.~\cite{BS11} is a mere toy model, only reproducing a single void in a FLRW background.
The important results of our paper are to show that, even if a QSS model of this kind exhibits a cosmological blueshift,
the averaging process transforms it into a cosmological redshift which is in accordance with observations and that the 
redshift-drift can, in principle, allow us to discriminate between the averaged model and the $\Lambda$CDM model while 
both reproduce the same cosmological data on the observer's past light cone.

It might happen that, in the future, more elaborate inhomogeneous models with no dark energy, such as Swiss-cheese 
models where the patches could be QSS without any symmetry and whose average might be LTB Swiss-cheeses reproducing 
the cosmological data, or QSS Swiss-cheese models reproducing themselves the data, should be proposed in the literature. In this case, our work could serve as a recipe to 
calculate the redshift and a then measurable redshift-drift in these models. It has been indeed shown in Ref.~\cite{JL08} that
a 42-m telescope is able of unambiguously detect the redshift-drift over a 20 year period at a redshift $2<z<5$. Therefore,
if one constructs a QSS Swiss-cheese model of the kind described above reaching a redshift of at least $z=2$,  the comparison  with measured
redshift-drifts  might become possible in  the future.



\end{document}